\documentclass[MSNbibl,nameyear,dvips]{arxstspdf}
\usepackage{flushend}
\usepackage{stfloats}
\usepackage{graphicx}
\volume{29}
\issue{1}
\pubyear{2014}
\firstpage{106}
\lastpage{112}
\doi{10.1214/13-STS417} 



\begin{document}
\begin{frontmatter}

\title{The Two-Piece Normal, Binormal, or Double Gaussian
Distribution: Its Origin and Rediscoveries}
\runtitle{The Two-Piece Normal Distribution}

\begin{aug}
\author[a]{\fnms{Kenneth F.} \snm{Wallis}\corref{}\ead[label=e1]{K.F.Wallis@warwick.ac.uk}}
\runauthor{K.~F. Wallis}

\affiliation{University of Warwick}

\address[a]{Kenneth F. Wallis is Emeritus Professor of Econometrics, Department of Economics, University of Warwick,
Coventry CV4 7AL, United Kingdom \printead{e1}.}

\end{aug}

\begin{abstract}
This paper traces the history of the two-piece normal
distribution from its origin in the posthumous
\textit{Kollektivmasslehre} (1897) of Gustav Theodor Fechner to its
rediscoveries and generalisations. The denial of Fechner's originality
by Karl Pearson, reiterated a century later by Oscar Sheynin, is shown
to be without foundation.
\end{abstract}


\end{frontmatter}

\section{Introduction}\label{sec1}

The two-piece normal distribution came to public attention in the late
1990s, when the Bank of England and the Sveriges Riksbank began to
publish probability forecasts of future inflation, using this
distribution to represent the possibility that the balance of risks
around the central forecast might not be symmetric. The forecast
probabilities that future inflation would fall in given intervals could
be conveniently calculated by scaling standard normal probabilities, and
the resulting density forecasts were visualised in the famous forecast
fan charts. In both cases the authors of the supporting technical
documentation (Britton, Fisher and Whitley, \citeyear{BRIFISWHI98}; \cite*{BLISEL})
refer readers to Johnson, Kotz and Balakrishnan (\citeyear{JohKotBal94}) for discussion of
the distribution. These last authors state (page~173) that ``the
distribution was originally introduced by \citet{GIBMYL73},'' a
reference that post-dates the first edition of \textit{Distributions in
Statistics} (\cite*{JOHKOT70}), in which the two-piece normal
distribution made no appearance, under this or any other name. On the
contrary, the distribution was originally introduced in Fechner's
\textit{Kollektivmasslehre} (\citeyear{FECN2}) as the \textit{Zweispaltiges} or
\textit{Zweiseitige Gauss'sche Gesetz}. In his monumental history of
statistics, \citet{Hal98} prefers the latter name, which translates as the
``two-sided Gaussian law,'' and refers to it as ``the Fechner distribution''
(page~378). However Fechner's claim to originality had been disputed by
\citet{PEA05}, whose denial of Fechner's originality has recently been
repeated by \citet{She04}. In this paper we reappraise the source and
nature of the various claims, and record several rediscoveries of the
distribution and extensions of Fechner's basic ideas. As a prelude to
the discussion, there follows a brief technical introduction to the
distribution.

\begin{figure}

\includegraphics{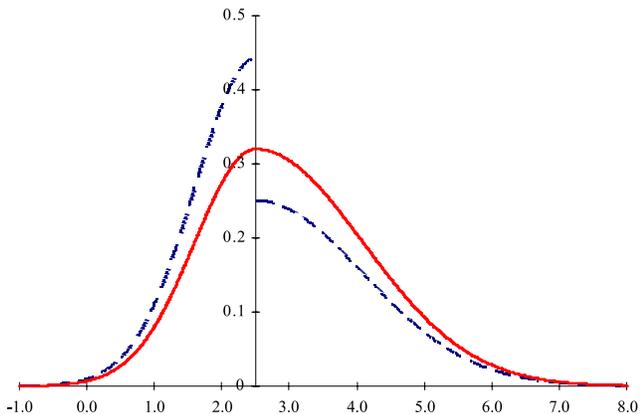}

\caption{The probability density function of the two-piece normal
distribution.
Dashed line: left half of $N(\mu,\sigma_{1})$ and right half of
$N(\mu,\sigma_{2})$ distributions with
$\mu = 2.5$ and $\sigma_{1} < \sigma_{2}$.
Solid line: the two-piece normal distribution.}\label{fig1}
\end{figure}

A random variable $X$ has a two-piece normal distribution with
parameters
$\mu, \sigma_{1}\mbox{ and }\sigma_{2}$ if it has probability
density function (PDF)
\begin{equation}
\label{eq1} f(x) = \cases{\displaystyle A\exp \bigl[ - ( x - \mu
)^{2} / 2\sigma_{1}^{2} \bigr],&\quad $x \le\mu,$
\vspace*{2pt}\cr
\displaystyle A\exp \bigl[ - ( x - \mu )^{2} / 2\sigma_{2}^{2}
\bigr],&\quad $x \ge\mu,$}
\end{equation}
where $A = ( \sqrt{2\pi} ( \sigma_{1} + \sigma_{2} ) / 2 )^{ - 1}$. The
distribution is formed by taking the left half of a normal distribution
with parameters $(\mu, \sigma_{1})$ and the right half of a
normal distribution with parameters $(\mu, \sigma_{2})$, and
scaling them to give the common value $f(\mu) = A$ at the mode, $\mu$,
as in (\ref{eq1}). The scaling factor applied to the left half of the $N(
\mu,\sigma_{1} )$ PDF is $2\sigma_{1} / ( \sigma_{1} + \sigma_{2} )$
while that applied to the right half of $N( \mu,\sigma_{2} )$ is
$2\sigma_{2} / ( \sigma_{1} + \sigma_{2} )$, so the probability mass
under the left or right piece is $\sigma_{1} / ( \sigma_{1} + \sigma_{2}
)$ or $\sigma_{2} / ( \sigma_{1} + \sigma_{2} )$, respectively. An
example with $\sigma_{1} < \sigma_{2}$, in which the two-piece normal
distribution is positively skewed, is shown in Figure \ref{fig1}. The skewness
becomes extreme as $\sigma_{1} \to 0$ and the distribution collapses to
the half-normal distribution, while the skewness is reduced as
$\sigma_{1} \to\sigma_{2}$, reaching zero when $\sigma_{1} = \sigma_{2}$
and the distribution is again the normal distribution.

The mean and variance of the distribution are
\begin{eqnarray}
\label{eq2} E(X) &=& \mu + \sqrt{\frac{2}{\pi}} ( \sigma_{2} -
\sigma_{1} ),
\\
\label{eq3} \operatorname{var} (X) &=& \biggl( 1 - \frac{2}{\pi} \biggr)
( \sigma_{2} - \sigma_{1} )^{2} +
\sigma_{1}\sigma_{2} .
\end{eqnarray}
Expressions for the third and fourth moments about the mean are
increasingly complicated and uninformative. Skewness is more readily
interpreted in terms of the ratio of the areas under the two pieces of
the PDF, which is $\sigma_{1} / \sigma_{2}$, or a monotone
transformation thereof such as $( \sigma_{2} - \sigma_{1} ) / (
\sigma_{2} + \sigma_{1} )$, which is the value taken by the skewness
measure of \citet{ArnGro95}. With only three parameters
there is a one-to-one relation between (the absolute value of) skewness
and kurtosis. The conventional moment-based measure of kurtosis,
$\beta_{2}$, ranges from 3 (symmetry) to 3.8692 (the half-normal extreme
asymmetry), hence the distribution is leptokurtic.

Quantiles of the distribution can be conveniently obtained by scaling
the appropriate standard normal quantiles. For the respective cumulative
distribution functions (CDFs) $F(x)$ and $\Phi(z)$ we define quantiles
$x_{p} = F^{ - 1}(p)$ and $z_{p} = \Phi^{ - 1}(p)$. Then in the left
piece of the distribution we have $x_{\alpha  } = \sigma_{1}z_{\beta  }
+ \mu$, where $\beta = \alpha( \sigma_{1} + \sigma_{2} ) / 2\sigma_{1}$.
And in the right piece of the distribution, defining quantiles with
reference to their upper tail probabilities, we have $x_{1 - \alpha  } =
\sigma_{2}z_{1 - \delta  } + \mu$, where $\delta = \alpha( \sigma_{1} +
\sigma_{2} ) / 2\sigma_{2}$. In particular, with $\sigma_{1} <
\sigma_{2}$, as in Figure \ref{fig1}, the median of the distribution is $x_{0.5}
= \sigma_{2}\Phi^{ - 1}( 1 - (\sigma_{1} + \sigma_{2}) / 4\sigma_{2} ) +
\mu$. In this case the three central values are ordered
mean$>$median$>$mode; with negative skewness this order is reversed.

Although the two-piece normal PDF is continuous at $\mu$, its first
derivative is not and the second derivative has a break at $\mu$, as
first noted by \citet{RANGRE04}. This has the disadvantage of
making standard asymptotic likelihood theory inapplicable, nevertheless
standard asymptotic results are available by direct proof for the
specific example.

The remainder of this paper is organised as follows. In Section \ref{sec2} we
revisit the distribution's origin in Gustav Theodor Fechner's
\textit{Kollektivmasslehre}, edited by Gottlob Friedrich Lipps and
published in 1897, ten years after Fechner's death. In Section \ref{sec3} we note
an early rediscovery, two years later, by Francis Ysidro Edgeworth. In
Section \ref{sec4} we turn to the first discussion in the English language of
Fechner's contribution, in a characteristically long and argumentative
article by Karl \citet{PEA05}. Pearson derives some properties of
``Fechner's double Gaussian curve,'' but asserts that it is ``historically
incorrect to attribute [it] to Fechner.'' We re-examine Pearson's
evidence in support of this position, in particular having in mind its
reappearance in Oscar Sheynin's (\citeyear{She04}) appraisal of Fechner's
statistical work. Pearson also argues that ``the curve is not general
enough,'' especially in comparison with his family of curves. The overall
result was that the Fechner distribution was overlooked for some time,
to the extent that there have been several independent rediscoveries of
the distribution in more recent years; these are noted in Section \ref{sec5},
together with some extensions.

\section{The Originators: Fechner and Lipps}\label{sec2}

Gustav Theodor Fechner (1801--1887) is known as the founder of
psychophysics, the study of the relation between psychological sensation
and physical stimulus, through his \citeyear{FECN1} book \textit{Elemente der
Psychophysik}. Stigler's (\citeyear{Sti86}, pages~242--254) assessment of this
``landmark'' contribution concludes that ``at a stroke, Fechner had created
a methodology for a new quantitative psychology.'' However, his final
work, \textit{Kollektivmasslehre}, is devoted more generally to the
study of mass phenomena and the search for empirical regularities
therein, with examples of frequency distributions taken from many
fields, including aesthetics, anthropology, astronomy, botany,
meteorology and zoology. In his Foreword, Fechner mentions the long
gestation period of the book, and states its main objective as the
establishment of a generalisation of the Gaussian law of random errors,
to overcome its limitations of symmetric probabilities and relatively
small positive and negative deviations from the arithmetic mean. He also
appeals to astronomical and statistical institutes to use their
mechanical calculation powers to produce accurate tables of the Gaussian
distribution, which he had desperately missed during his work on the
book. But the book had not been completed when Fechner died in November
1887.

The eventual publication of \textit{Kollektivmasslehre} in 1897 followed
extensive work on the incomplete manuscript by Gottlob Friedrich Lipps
(1865--1931). In his Editor's Preface, Lipps says that he received the
manuscript in early 1895 and that material he has worked on is placed in
square brackets in the published work. It is not clear how much
unfinished material was left behind by Fechner or to what extent Lipps
had to guess at Fechner's intentions. It would appear that the overall
structure of the book had already been set out by Fechner, since most of
the later chapters have early paragraphs by Fechner, before
square-bracketed paragraphs begin to appear. Also, some earlier chapters
by Fechner have forward references to later material that appears in
square brackets. In general, Lipps' material is more mathematical: he
was more of a mathematician than Fechner, who perhaps had set some
sections aside for attention later, only to run out of time. Lipps also
has a lighter style: for example, Sheynin (\citeyear{She04}, page~54) complains
about some earlier work that ``Fechner's style is troublesome. Very often
his sentences occupy eight lines, and sometimes much more---sentences
of up to 16 lines are easy to find.'' The same is true of the present
work.

The origin of the two-piece normal distribution is in Chapter 5 of
\textit{Kollektivmasslehre}, titled ``The Gaussian law of random
deviations and its generalisations.'' Here Fechner uses very little
mathematics, postponing more analytical treatment to later chapters. He
first presents a numerical example of the use of the Gaussian
distribution to calculate the probability of an observation falling in a
given interval. The measure of location is the arithmetic mean, $A$, and
the measure of dispersion is the mean absolute deviation, $\varepsilon$
(related to the standard deviation, in the Gaussian distribution, by
$\varepsilon = \sigma\sqrt{2 / \pi} )$. Tables of the standard normal
distribution are not yet available, and his calculations proceed via the
error function (see \cite*{Sti86}, pages~246--248, e.g.), and prove
to be remarkably accurate.

In previous work Fechner had introduced other ``main values'' of a
frequency distribution, the \textit{Zentralwert} or ``central value'' $C$,
and the \textit{Dichteste Wert} or ``densest value'' $D$, subsequently
known in English as the median and the mode. Arguing that the equality
of $A$, $C$ and $D$ is the exception rather than the rule, he next
introduces the \textit{Zweispaltiges} \textit{Gauss'sche Gesetz} to
represent this asymmetry. Calculating mean absolute deviations from the
mode separately for positive and negative deviations from $D$, the ``law
of proportions'' is invoked, that these should be in the same ratio as
the numbers of observations on which they are based. On converting from
relative frequencies of observations to probabilities, and from subset
mean absolute deviations to subset standard deviations, it is seen that
this is exactly the requirement discussed above, that the probabilities
below and above the mode are in the ratio $\sigma_{1} / \sigma_{2}$, to
give a curve that is continuous at the mode. Fechner says that he first
discovered this law empirically, and warns that determination of the
mode from raw data is not straightforward. He goes on to show that, in
this distribution, the median lies between the mean and the mode.

The first mathematical expression of the two normal curves with
different precision soon appears in what is the first square-bracketed
paragraph in the book and the only such paragraph in Chapter 5. More
extensive workings by Lipps appear in Chapter 19, ``The asymmetry laws,''
where every paragraph is enclosed in square brackets. Here Lipps traces
the development and properties of the distribution more formally,
including an expression for the density function [equation (6),
page~297] which corresponds to equation (\ref{eq1}) on converting between
measures of dispersion. Nevertheless, the key steps in that development,
in Chapter~5, were Fechner's alone.

We note that the second ``generalisation'' presented later in Chapter 5 of
\textit{Kollektivmasslehre} is a form of log-normal distribution, but
this receives less emphasis and is not our present focus of attention.

\section{An Early Rediscovery: Edgeworth}\label{sec3}

In 1898--1900 Edgeworth contributed a five-part article ``On the
representation of statistics by mathematical formulae'' to the
\textit{Journal of the Royal Statistical Society}, each part appearing
in a different issue of the journal. His objective was ``to recommend
formulae which have some affinity to the normal law of error, as being
specially suited to represent statistics of frequency.'' The first two
parts deal with the ``method of translation,'' or transformations to
normality, and the ``method of separation,'' or mixtures of normals, using
modern terminology.

In the third part Edgeworth considers the ``method of composition,'' in
which he constructs ``a \textit{composite probability-curve}, consisting
of two half-probability curves of different types, tacked together at
the \textit{mode}, or greatest ordinate, of each, so as to form a
continuous whole, as in the accompanying figure'' (1899, page~373,
emphasis in original; the figure is very similar to the solid line in
Figure \ref{fig1} above). He gives expressions for the two appropriately scaled
half-normal curves, as above, using the \textit{modulus}, equal to
$\sqrt{2}$ standard deviation, as his preferred measure of spread. He
says that this idea of two probability curves with different moduli is
suggested by \citet{LUD98}; however, its development in the context of
the normal distribution is Edgeworth's alone, since Ludwig's comment
comes in a discussion of frequency curves based on the binomial
distribution.

To ``determine the constants,'' that is, estimate the parameters, given a
sample mean and second and third sample moments, Edgeworth rearranges
their definitions to obtain a cubic equation in the distance between the
mean and the mode; the required parameter estimates follow from the real
solution to this equation. He gives a practical example and compares the
method of composition to the methods discussed earlier. In his opinion,
the ``essential attribute'' of the new method is its ``deficiency of
{a priori} justification,'' in contrast to the normal distribution
itself.

\section{The Critics: Pearson and Sheynin}\label{sec4}

The first English-language discussion of Fechner's contribution appears
in a 44-page article by Karl Pearson, published in 1905 in
\textit{Biometrika},\vadjust{\goodbreak} the journal he had co-founded four years earlier.
The article is a response to a review of Pearson's and Fechner's works
on skew variation by \citet{RANGRE04} in the leading German
anthropology journal. Pearson's title quotes most of the title of the
German article, omitting its reference to anthropology, and adds the
words ``A rejoinder,'' although the running head throughout his article is
``Skew variation, a rejoinder.'' He explains that the German journal had
provisionally accepted a rejoinder, but when it arrived the editors did
not ``see fit to publish'' his reply, so he placed it in
\textit{Biometrika}, of which he was, in effect, managing editor. From a
statistical point of view this seems to have been a more appropriate
outcome, since his article contains much general statistical discussion
and is most often cited for its introduction of the terms platykurtic,
leptokurtic and mesokurtic.

However, Pearson's article also contains extensive attacks on Ranke and
Greiner, who had argued that, for the anthropologist, only the Gaussian
law is of importance. In this respect the article is a good example of
his well-documented behaviour. For example, Stigler (\citeyear{Sti99}, Chapter~1)
opens by observing that ``Karl Pearson's long life was punctuated by
controversies, controversies he often instigated, usually pursued with a
zealous energy bordering on obsession;'' he ``was a fighter who vigorously
reacted against opinions that seemed to detract from his own theories.
Instead of giving room for other methods and seeking cooperation, his
aggressive style led to controversy'' (\cite*{Hal98}, page~651); he was ever
``relentless in controversy'' (\cite*{Cox01}, page~5) and ``beyond question a
fierce antagonist'' (\cite*{Por04}, page~266). Some of this antagonism is
directed towards Fechner: although Pearson and Fechner are on the same
side of the debate with Ranke and Greiner about asymmetry, Pearson sees
``Fechner's double Gaussian curve'' as a rival to his family of curves,
and criticises it on both statistical and historical grounds.

Using the parameterisation in terms of $\sigma_{1}$ and $\sigma_{2}$ as
in equation (\ref{eq1}), Pearson presents expressions for the first four moments
of the distribution. Rather than ``the rough process by which Fechner
determines the mode and obtains the constants of the distribution,'' he
shows that ``fitting by my method of moments is perfectly
straightforward.'' To do this, he obtains the cubic equation discussed
above, and says in a footnote (page~197) ``This cubic was, I believe,
first given by Edgeworth,'' but there is no reference.\vadjust{\goodbreak} He observes that
the skewness and kurtosis are not independent of one another, so that
``we cannot have any form of symmetry but the mesokurtic.'' He obtains the
bounds on $\beta_{2}$ given above, but notes that many empirical
distributions with values outside this range have been observed. Hence,
Pearson's overall conclusion is that ``the double Gaussian curve fails us
hopelessly.'' Curiously, having defined platykurtic as ``more flat-topped''
and leptokurtic as ``less flat-topped'' than the normal curve, as has
become standard usage, he contrarily describes Fechner's double Gaussian
curve as platykurtic, despite having shown its positive excess kurtosis.
Similarly, another curve, the symmetrical binomial, is said to be
``essentially leptokurtic, that is, $\beta_{2} < 3$'' (page~175).

Turning to questions of precedence, Pearson's counter claims appear in a
footnote (page~196) at the start of the statistical discussion summarised
above, which reads as follows:
\begin{quote}
Here again it is historically incorrect to attribute these curves to
Fechner. They had
been proposed by De Vries in 1894, and termed ``half-Galton curves,''
and Galton was
certainly using them in 1897. See the discussion in Yule's memoir,
\textit{R. Statist. Soc.}
\textit{Jour.} Vol. LX, page~45 \textit{et seq.}
\end{quote}

Pearson was familiar with \citet{DE1894}, having used two of his
J-shaped botanical frequency distributions as Examples XI and XII in his
\citeyear{PEA} article on skew variation. De Vries said that these deserved the
name half-Galton (i.e., half-normal) simply on the basis of the
appearance of the empirical distributions, and no fitting was attempted,
nor did he make any proposal to place two such curves together to give a
more general asymmetric distribution. Fechner's curve had \textit{not}
been proposed by De Vries. [Edgeworth knew that his composite curve had
not either, noting at the outset (\citeyear{EDG99}, page~373) that ``It will be
observed that the following construction is not much indebted to the
``half-Galtonian'' curve employed by Professor De Vries.'']

Galton comes a little closer, but Pearson is again incorrect. His
citation is inaccurate, since he clearly has in mind Yule's paper read
at the Royal Statistical Society in January 1896, published with
discussion later that year (\cite*{YUL96N1}). Galton had opened the
discussion at the meeting and mentioned his method of percentiles as an
alternative to the method of frequency curves developed by Pearson and
applied by Yule. In response to a request at the meeting, he provided a
memorandum giving\vadjust{\goodbreak} fuller information on his method, which was published
in the same issue of the Society's journal (\cite*{GAL96}), together with
a reply by \citet{YUL96N2}. Galton explains how his method of percentiles,
in this example method of deciles, smooths the original frequency table
or ``frequency polygon'' of Yule by interpolating deciles and plotting
them. He then mentions another approach, namely

\begin{quote}
$\ldots$ the extremely rude and scarcely defensible method, but still a
sometimes
serviceable one, of looking upon skew-curves as made up of the halves of
two
different normal curves pieced together at the mode. $\ldots$ On trying
it, again for
curiosity's sake, with the present series for all the five years, there
was of course
no error for the 2nd, 5th, and 8th deciles, $\ldots$
\end{quote}
because he had inferred the spread or standard deviation of the lower
half-normal distribution from the lower 20\% point of the standard
normal distribution, and similarly for the upper part; he goes on to
discuss the errors of fit at the other deciles. But no ``law of
proportions'' or scaling is applied, and the resulting curve is
discontinuous, like the initial two halves of normal curves in Figure \ref{fig1}.
\citet{YUL96N2} recognises this in his response, noting that, in contrast,
his skew-curve ``presents a continuous distribution round the mode.''
Galton was certainly \textit{not} using Fechner's curves.

The erroneous assertions in Pearson's footnote may be due to his
combativeness. Several authors also discuss the tremendous volume of
work he undertook. For example, Cox (\citeyear{Cox01}, page~6) observes that he
``wrote more than 90 papers in \textit{Biometrika} in the period up to
1915, few of them brief, and appears to have been the moving spirit
behind many more.'' He founded not only the journal but also the
Biometrics Laboratory at University College London at this time. His son
Egon remarks that the volume of work ``led inevitably to a certain hurry
in execution'' (E.~S. \cite*{PEA36}, page~222). This remark is made during
discussion of one of Pearson's two well-known errors, recently
reappraised by \citet{Sti08}, but it perhaps also applies to the
mistakes discussed above, which are of a smaller order of magnitude.
Nevertheless, Pearson's assertions in the quoted footnote are mistaken,
and his challenge to Fechner's claim to priority is unjustified, and
thereby unjust.

\citet{She04}, in his review of Fechner's statistical work, has a very
brief discussion of the\vadjust{\goodbreak} double-sided Gaussian law, quoting from sections
of \textit{Kollektivmasslehre} that had been worked on by Lipps, and
hence underestimating the role of Fechner's law of proportions. In his
discussion (page~68) he states that the double-sided Gaussian law was not
original to Fechner, this having been pointed out, forcefully, by
\citet{PEA05}. As if quoting from Pearson, and giving no citation for
\citet{DE1894}, Sheynin states ``De Vries, in 1894, had applied the
double-sided law.'' In this statement ``applied'' is somewhat stronger than
Pearson's ``proposed,'' hence is further from the truth, and Sheynin's
denial of Fechner's originality is similarly inaccurate and unjust.

\section{Later Rediscoveries and Extensions}\label{sec5}

The result of Pearson's critique appears to have been that, with two
exceptions discussed below, the Fechner distribution, with this
attribution, disappeared from the statistical literature until its
reappearance in Hald's (\citeyear{Hal98}) history. Meanwhile, three independent
rediscoveries occurred.

First, in the physics literature, is Gibbons and Mylroie's (\citeyear{GIBMYL73})
``joined half-Gaussian'' distribution, cited by Johnson, Kotz and
Balakrishnan (\citeyear{JohKotBal94}), as noted above; the distribution is fitted by what
statisticians recognise as the method of moments. Second, in the
statistics literature, is the ``three-parameter two-piece normal''
distribution of \citet{Joh82}, also cited by Johnson, Kotz and
Balakrishnan~(\citeyear{JohKotBal94}); John compares estimation by the method of moments and maximum
likelihood. In the same journal \citet{Kim85} notes that \citet{Joh82} is
a rediscovery, with reference to \citet{GIBMYL73}; he proves
the asymptotic normality of ML estimators and provides a likelihood
ratio test of symmetry. Finally, in the meteorology literature, \citet{TOTSZE90}
introduce the ``binormal'' distribution, again fitted by
ML, with a test of symmetry. The same name is used by \citet{GARMcC97},
who nevertheless again attribute the distribution to Gibbons and
Mylroie. In all these articles the distribution is parameterised in
terms of the mode, using various symbols, and the standard deviations
$\sigma_{1}$ and $\sigma_{2}$, as in (1) above. An alternative
parameterisation, with a single explicit skewness parameter, is given by
\citet{MudHut00}, who do acknowledge Fechner's priority.

A modern, but pre-\citet{Hal98} attribution to \textit{Kollektivmasslehre}
occurs at the start of an exploration by \citet{Run78} of the mean,
median, mode ordering. He notes that Fechner had shown this
\textit{Lagegesetz der Mittelwerte} for the two-piece normal
distribution, and investigates more general conditions in which it
holds. The second exceptional appearance of the Fechner distribution in
the statistical literature pre-1998 is more substantial. \citet{BAR},
seeking a family of distributions ``which may be expected to represent
most of the types of skewness liable to arise in practice,'' introduces
the distribution
\[
f(x) = \cases{\displaystyle K\exp \biggl[ - \frac{1}{2} \biggl(
\frac{ -
M ( x - \mu  )}{\sigma} \biggr)^{a} \biggr],&\quad $x \le\mu,$
\vspace*{2pt}\cr
\displaystyle K\exp \biggl[ - \frac{1}{2} \biggl( \frac{x - \mu} {\sigma}
\biggr)^{a} \biggr],&\quad $x \ge\mu,$}
\]
which reparameterises and generalises the two-piece normal distribution
in equation (\ref{eq1}). He calls it the Fechner family, because by allowing the
skewness parameter $M\ ( M > 0 )$ to differ from 1 it embodies Fechner's
idea in \textit{Kollektivmasslehre} of having different scales for
positive and negative deviations from the mode, $\mu$. It also allows
for nonnormal kurtosis by allowing $a\ ( 1 \le a < \infty)$ to differ
from 2. The scale parameter $\sigma$ is equal to the standard deviation
if $( M,a ) = ( 1,2 )$ but not otherwise, in general. This ``Fechner
family of unimodal densities'' also appears in a later article (\cite*{BAR95}),
which is cited by Hald (\citeyear{Hal98}, page~380). We note that the case $a
= 1$, the asymmetric Laplace distribution, has a considerable life of
its own, beginning before Barnard's work: see, for example, Kotz,
Kozubowski and Podgorski (\citeyear{KotKozPod01}, Chapter~3) and the references therein.

Two further extensions of note, independent of Fechner, can be found in
Bayesian statistics. For the application of Monte Carlo integration with
importance sampling to Bayesian inference, \citet{Gew89} uses ``split''
(i.e., two-piece) multivariate normal and Student-$t$ distributions
as importance sampling densities. The generalisation by Fernandez and
Steel (\citeyear{FerSte98}) is also cast in a Bayesian setting: as in Barnard's Fechner
family, there is a single skewness parameter, which is convenient
whenever it is desired to assign priors to skewness; nevertheless, it
has general applicability. For any univariate PDF $f(x)$ which is
unimodal and symmetric around 0, Fernandez and Steel's class of
two-piece or split distributions $p( x |\gamma)$, indexed by a skewness
parameter $\gamma\ ( \gamma > 0 )$, is
\begin{eqnarray}
p( x |\gamma) = \cases{\displaystyle K f( \gamma x ),&\quad $x \le 0,$ \vspace*{2pt}\cr
\displaystyle K f\biggl(
\frac{x}{\gamma} \biggr),&\quad $x \ge 0$, }
\end{eqnarray}
where $K = 2\bigl( \gamma
+ \gamma^{ - 1} \bigr)^{ - 1}$.
If $\gamma > 1$ there is positive skewness, and inverting $\gamma$
produces the mirror image of the density function around 0. Unlike
Barnard's Fechner family there is no explicit kurtosis parameter;
kurtosis is introduced, if desired, by the choice of $f(x)$, most
commonly as Student-$t$. An extension with two tail parameters to allow
different tail behaviour in an asymmetric two-piece $t$-distribution is
developed by \citet{ZhuGal10}.

\section*{Acknowledgments}
I am grateful to Sascha Becker, David Cox,
Chris Jones, Malte Kn\"{u}ppel, Kevin McConway, Mary Morgan, Oscar
Sheynin, Mark Steel, Stephen Stigler and two referees for comments and
suggestions at various stages of this work. I am especially indebted to
Karl-Heinz T\"{o}dter for his expert assistance with the German-language
works cited. Thanks also to the Resource Delivery team at the University
of Warwick Library.



\end{document}